\documentstyle[psfig,12pt]{article}

\def\refe{\par\noindent\hangindent=1.5cm}
\begin{document}
\begin{center}
{\large \bf The Temporal Aspect of the Drake Equation and SETI}

\vspace{0.5cm}

\large Milan M.\ \'Cirkovi\'c

\vspace{0.1cm}

{\it Astronomical Observatory Belgrade, \\
Volgina 7, 11160 Belgrade, Serbia and Montenegro \\

email: {\tt mcirkovic@aob.aob.bg.ac.yu} }

\vspace{0.7cm}
\end{center}

\begin{abstract}
We critically investigate some evolutionary aspects of the famous
Drake equation, which is usually presented as the central guide
for the research on extraterrestrial intelligence. It is shown
that the Drake equation tacitly relies on unverifiable and
possibly false assumptions on both the physico-chemical history
of our Galaxy and the properties of advanced intelligent
communities. The importance of recent results of Lineweaver on
chemical build-up of inhabitable planets for SETI is emphasized.
Two important evolutionary effects are briefly discussed and the
resolution of the difficulties within the context of the
phase-transition astrobiological models sketched.
\end{abstract}


\section{Introduction}
It is hard to deny that the Search for ExtraTerrestrial
Intelligence (SETI) is one of the major scientific adventures in
the history of humankind. At the beginning of XXI century it
remains the oldest and perhaps the most fascinating scientific
problem. However, the field is still largely qualitative and thus
often not taken seriously enough. One of the attempts to overcome
this circumstance is encapsulated in the famous Drake equation,
developed by Frank Drake for the first SETI symposium in 1961.

The first problem any student of SETI faces is that there is no
canonical form of the Drake equation. Various authors quote
various forms of the equation, and it is in a sense dependent on
what is the desired result of the analysis. We shall investigate
the following form (e.g.\ Shklovskii and Sagan 1966; Walters,
Hoover, and Kotra 1980):

\begin{equation}
\label{drake} N = R_\ast f_g f_p n_e f_l f_i f_c L ,
\end{equation}
while keeping in mind that other equivalent forms exist as well.
In this expression, the symbols have the following meanings:

\vspace{0.2cm}

$N$ = the number of Galactic civilizations with whom communication
is possible.

$R_\ast$ = mean rate of star formation in the Galaxy,

$f_g$ = fraction of stars suitable for supporting life,

$f_p$ = fraction of stars with planetary systems,

$n_e$ = number of planets per planetary system with conditions
ecologically suitable for the origin and evolution of life,

$f_l$ = fraction of suitable planets where life originates and
evolves into more complex forms,

$f_i$ = fraction of planets bearing life with intelligence,

$f_c$ = fraction of planets with intelligence that develops a
technological phase during which there is the capability for an
interest in interstellar communication,

$L$ = mean lifetime of a technological civilization.

\vspace{0.2cm}

Almost all authors agree on the general meanings of various
$f$-parameters and $n_e$ (though wildly differing in the values
they ascribe to each of them!); on the other hand, the product
$R_\ast L$ is sometimes written in the form
\begin{equation}
\label{dva} R_\ast L = n_\ast \frac{L}{t_0},
\end{equation}
where $n_\ast$ is the {\it current\/} number of stars in the
Galaxy, and $t_0$ is the age of our stellar system (currently
thought to be $t_0 \approx 12$ Gyr; e.g.\ Krauss and Chaboyer
2003). This is useful since (i) $R_\ast$ is not a directly
measurable quantity while $n_\ast$ and $t_0$ are, at least in
principle, and (ii) it enables direct comparison of two
characteristic timescales, cosmological ($t_0$) and
"astrosociological" ($L$). There is a catch in (\ref{dva}),
however, since the star formation rate is not uniform throughout
the history of the Galaxy, and thus in general $\langle R_\ast
\rangle \neq n_\ast / t_0$. While this particular problem is not
acute from the SETI point of view, due to the metallicity effects
(early epochs of intense star formation are characterized by low
metallicity), it points in the direction of similar difficulties
following from unwarranted assumptions of uniformity. We argue
below that the main shortcoming of the Drake equation is its lack
of temporal structure, i.e., it fails to take into account various
evolutionary processes which form a pre-requisite for anything
quantified by $f$-factors and $n_e$.

It is important to understand that we are criticizing the Drake
equation not as an expression {\it per se}, but as a guiding line
for a rather specific set of programs, procedures and (in the
final analysis) investments, known overall as SETI. In SETI
research proposals, Eq.\ (\ref{drake}) figures very prominently.
Both supporters and opponents of SETI invoke the same simple
numerical relationship in order to promote their respective
views. However, arguments for both side are to be seriously
doubted if the underlying relationship has serious deficiencies
for any practical application, e.g.\ for estimating the timescale
for sustained SETI effort by which we might expect to detect
extraterrestrial intelligent signals (or artifacts).

Probably the most telling argument as to the inadequacy of the
Eq.~(1) is that almost four decades of SETI projects have not
given results, in spite of the prevailing "contact optimism" of
1960s and 1970s, motivated largely by uncritical acceptance of
the Drake equation. Conventional estimates of that period spoke
about $10^6 - 10^9$ (!) advanced societies in the Milky Way
forming the "Galactic Club" (Bracewell 1975). Nowadays, even the
greatest SETI optimists have abandoned such fanciful numbers, and
settled to a view that advanced extraterrestrial societies are
much rarer than previously thought. One of the important factors
in this downsizing of SETI expectations has certainly been
demonstrations by "contact pessimists", especially Michael Hart
and Frank Tipler, that the colonization---or at least visit---of
all stellar systems in the Milky Way by means of self-reproducing
von Neumann probes is feasible within a minuscule fraction of the
Galactic age (Hart 1975; Tipler 1980). In this light, Fermi's
legendary question: {\it Where are they?\/} becomes disturbingly
pertinent. In addition, Carter (1983) suggested an independent
and powerful anthropic argument for the uniqueness of intelligent
life on Earth in the Galactic context. Today, it is generally
recognized that "contact pessimists" have a strong position. How
then, one is tempted to ask, does the discrepancy with our best
analyses of Eq.~(\ref{drake}) arise?

 Now we shall show that there are
two serious problems which make the equation (\ref{drake}) much
less practical from the SETI point of view than has been
conventionally thought. The two have the opposite effect on $N$,
and may well partially cancel one another out; still, by a careful
consideration those effects could be disentangled. Some other
criticisms of the Drake equation, from different points of view,
can be found in Walters et al.\ (1980), Wilson (1984), Ward and
Brownlee (2000), and Walker and \'Cirkovi\'c (2003, preprint).

\section{Upper limit on civilization's age}

In principle, the parameter $L$ in Eq.\ (\ref{drake}) could be
arbitrarily large, thus offsetting any exceptionally small value
among different $f$-parameters. Historically, that was the
conventional assumption of "contact optimists" like Sagan,
Shklovskii, or Drake in the earlier decades (1960s and 1970s) of
SETI efforts. It is reasonable to assume that after a
technological civilization overcomes its "childhood troubles"
(like the threat of destruction in a nuclear war or through the
misuse of nanotechnology) and starts colonizing space, it has
very bright prospects for survival on timescales of millions (if
not billions) years. Since it was intuitively clear (although
quantified only recently; see below) that most of the inhabitable
planets in the Milky Way are older than Earth, it was
hypothesized that civilizations to be found through SETI projects
will be significantly older than our civilization. However, it is
a leap of faith from a reasonable estimate of the temporal
distribution of civilizations to the assumption that we would be
able to communicate with them, or that they would express any
interest in communicating with us using our primitive
communication means. Even worse, people of the "contact optimism"
camp have been expressing hope that we would be able to {\it
intercept\/} communications between such very advanced societies,
which seems still less plausible.\footnote{For a profound and
poignantly ironic literary account of these issues see Lem (1984,
1987).}

Obviously, from (\ref{drake}) we have $\lim_{L \rightarrow
\infty} N = \infty$, which is senseless, for the finite spatial
{\it and\/} temporal region of spacetime we are considering in
practical SETI. And still, remarkably, it is {\it not\/} senseless
to contemplate upon the possibility that very advanced
civilizations can exist indefinitely in an open universe (e.g.\
Dyson 1979), i.e.\ that the limit $L \rightarrow \infty$ makes
sense. Whether an advanced technological society can exist
indefinitely---in accordance with the so-called Final Anthropic
Principle of Barrow and Tipler (1986) or the Final Anthropic
Hypothesis of \'Cirkovi\'c and Bostrom (2000)---is still an open
question in the nascent astrophysical discipline of {\it physical
eschatology\/} (Adams and Laughlin 1997; \'Cirkovi\'c 2003). Any
results from it, albeit very exciting and interesting in their own
right, are unimportant to SETI due to the large disparity of the
timescales involved.

According to a recent study by Lineweaver (2001), Earth-like
planets around other stars in the Galactic habitable zone are, on
average, 1.8 ($\pm 0.9$) Gyr older than our planet. His
calculations are based on chemical enrichment as the basic
precondition for the existence of terrestrial planets. Applying
the Copernican assumption naively, we would expect that
correspondingly complex lifeforms on those others to be, {\it on
average}, 1.8 Gyr older. Intelligent societies, therefore, should
also be older than ours by the same amount. In fact, the situation
is even worse, since this is just the average value, and it is
reasonable to assume that there will be, somewhere in the Galaxy,
an inhabitable planet (say) 3 Gyr older than Earth. Since the set
of intelligent societies is likely to be dominated by the small
number of oldest and most advanced members (for an ingenious
discussion in somewhat different context, see Olum 2003), we are
likely to encounter a civilization actually more ancient than 1.8
Gyr (and probably significantly more).

It seems preposterous to even remotely contemplate any possibility
of communication between us and Gyr-older supercivilizations. It
is enough to remember that 1 Gyr ago, the appearance of even the
simplest multicellular creatures on Earth lay in the distant
future. Thus, the set of the civilizations interesting from the
point of view of SETI is not open in the temporal sense, but
instead forms a "communication window", which begins at the
moment the required technology is developed (factor $f_c$ in the
Drake equation), and is terminated {\it either\/} through
extinction of the civilization {\it or\/} through it passing into
the realm of "supercivilizations" unreachable by our primitive
SETI means. Formally, this could be quantified by adding a term
to the Drake equation corresponding to the ratio of the duration
of the "communication window" and $L$. Let us call this ratio
$\xi$; we are, thus, justified in substituting $L$ in Eq.\
(\ref{drake}) with $\xi L$. Since $\xi$ is by definition smaller
than unity (and perhaps much smaller, if the present human
advances in communication are taken as a yardstick), the net
effect would be to drastically reduce the value of $N$.
Fortunately (from the SETI point of view) this is not the only
evolutionary effect hidden in the Drake equation.

\section{Simplicity of uniformitarianism}

A still more important shortcoming of Eq.~(\ref{drake}) as a
guideline to SETI consists of its uniform treatment of the
physical and chemical {\it history\/} of our Galaxy. It is
tacitly assumed that the history of the Galaxy is uniform with
respect to the emergence and capacities of technological
societies. This is particularly clear from the form (\ref{dva}),
as mentioned above. If, on the contrary, we assume more or less
sharply bounded temporal phases of the Galactic history as far as
individual terms in Eq.\ (\ref{drake}) are concerned, and take
into account our own existence at this particular epoch of this
history, we are likely to significantly underestimate the value
of $N$. We shall consider such a toy model below.

Uniformitarianism has not shone as a brilliant guiding principle
in astrophysics and cosmology. It is well-known, for instance,
how the strictly uniformitarian (and from many points of view
methodologically superior) steady-state theory of the universe of
Bondi and Gold (1948) and Hoyle (1948) has, after the "great
controversy" of 1950s and early 1960s, succumbed to the rival
evolutionary models, now known as the standard ("Big Bang")
cosmology (Kragh 1996). Balashov (1994) has especially stressed
this aspect of the controversy by showing how deeply justified
was the introduction of events and epochs never seen or
experienced by the Big Bang cosmologists. Similar arguments are
applicable in the nascent discipline of astrobiology, which might
be considered to be in an analogous state today as cosmology was
half a century ago.

The arguments of Lineweaver (2001) are crucial in this regard,
too. Obviously, the history of the Galaxy divides into at least
two periods (or phases): before and after sufficient metallicity
for the formation of Earth-like planets has been built up by
global chemical evolution. But this reflects only the most
fundamental division. It is entirely plausible that the history
of the Galaxy is divided still finer into several distinct
periods with radically different conditions for life. In that
case, only {\it weighted relative durations\/} are relevant, not
the overall age.

Exactly such a picture is presented by a class of phase
transition models (Clarke 1981; Annis 1999; see also Norris 2000),
which assume a {\it global regulation mechanism\/} for preventing
the formation of complex life forms and technological societies
early in the history of the Galaxy. Such a global mechanism could
have the physical form of $\gamma$-ray bursts, if it can be shown
that they exhibit sufficient lethality to cause mass biological
extinctions over a large part of the volume of the Galactic
habitable zone (Scalo and Wheeler 2002). If, as maintained in
these models, {\it continuous habitability\/} is just a myth, the
validity of the Drake equation (and the spirit in which it was
constructed and used) is seriously undermined.

For illustration, let us assume that the parameter $f_l$ has the
following evolutionary behavior:
\begin{equation}
\label{tri} f_l = \left\{ \begin{array}{r@{\quad:\quad}l}
10^{-6} &  0<t \leq t_p \\
0.9 &  t_p < t \leq t_0  \end{array} \right. .
\end{equation}
(we put the zero of time at the epoch of the Milky Way formation).
Here, $t_p$ is the epoch of global "phase transition" (Annis
1999), i.e.\ the epoch in which the lethal Galactic processes
became rare enough for sufficiently complex lifeforms to emerge.
Let us take $t_0 =12$ Gyr and $t_p = 11$ Gyr. Naive
uniformitarian application of the Drake equation would require us
to find the average $\langle f_l \rangle$ in particular example
$\langle f_l \rangle = 0.072$; if we assume $n_e =1$, other
$f$-factors all equal to 0.1 (rather conservative assumption), and
$R_\ast = 5$ yr$^{-1}$, we obtain that $N = 3.6 \times 10^{-5}
\xi L$, where $L$ is measured in years, and $\xi$ is the relative
duration of the communication window discussed above. In fact,
the true result is rather $N= 4.5 \times 10^{-4} \xi L$, more
than an order of magnitude higher. Such a big difference is of
obvious relevance to SETI; if $\xi L \sim 10^5$ yrs or less, it
might as well be the difference between sense and nonsense in the
entire endeavor. The discrepancy increases if the epoch of the
phase transition moves closer to the present time. The latter is
desirable if one wishes to efficiently resolve Fermi's "paradox"
through phase-transition models.

More realistically, we would expect several of the $f$-factors, as
well as $n_e$, to exhibit secular increase during the course of
Galactic history in a more complicated manner to be elaborated by
future detailed astrobiological models. Yet, steps similar to the
one in (\ref{tri}) seem inescapable at some point if we wish to
retain the essence of the phase transition idea. Barring this,
the only fully consistent and meaningful idea for both
explanation of the "Great Silence" and retaining the Copernican
assumption on Earth's non-special position is the "Interdict
Hypothesis" of Fogg (1987), as the generalized "Zoo Hypothesis"
(Ball 1973), which still seems inferior, since it explicitly
invokes non-physical (e.g., sociological) elements.

Intuitively, it is clear that in such phase transition models it
is a very sensible policy for humanity to engage in serious SETI
efforts: we expect practically all ETI societies to be roughly of
the same age as ours, and to be our competitors for
Hart-Tiplerian colonization of the Milky Way. The price to be paid
for bringing the arguments of "optimists" and "pessimists" into
accord is, obviously, the assumption that we are living in a
rather special epoch in Galactic history---i.e.\ the epoch of
"phase transition". That such an assumption is entirely
justifiable in the astrobiological context will be argued in a
subsequent study. (Parenthetically, this is entirely in accord
with the tenets of the currently much-discussed "rare Earth"
hypothesis; see Ward and Brownlee 2000.)

Note that in this case, the overall average age of a civilization
($L$) would give an entirely false picture at the outcome of the
Drake equation. In the toy model above, any hypothetical
civilization age of (say) 10 Gyr is obviously irrelevant
(although possibly sociologically allowed). This conclusion is
valid even if the width of the communication window is very
large, and in fact spans most of the lifetime of a civilization,
as SETI pioneers claimed ($\xi \approx 1$). Thus, we obtain a
physically more desirable picture for the explanation of Fermi's
"paradox" in which sociological influences are much less relevant.

\section{Conclusions}

We conclude that the Drake equation, as conventionally presented,
is not the best guide for both operational SETI and future
policy-making in this field. The reason for this is its lack of
temporal structure and appreciation of the importance of
evolutionary effects, so pertinent in the modern astrobiological
discourse. If we wish to go beyond the "zeroth-order"
approximation encapsulated by (\ref{drake}), we will need to
account for evolutionary effects, such as metallicity build-up
and "catastrophic" regulation of habitability. Notably, the
non-uniform history of the Galaxy as conceived in the phase
transition models can accommodate both the arguments of "contact
pessimists" and the justification for SETI projects, which have
been deemed incompatible in the literature so far. Future
detailed modeling will show in which way we can best accommodate
our knowledge on the history of the Galaxy in the overall
astrobiological picture.

\vspace{0.4cm}

\noindent {\bf Acknowledgements.} The author wholeheartedly thanks
Alan Robertson, Sa\v sa Nedeljkovi\'c, Vesna Milo\v
sevi\'c-Zdjelar, Srdjan Samurovi\'c, and Mark A. Walker for
invaluable technical help. Useful discussions with Nick Bostrom,
Robert J. Bradbury, Zoran \v Zivkovi\'c, Ivana Dragi\'cevi\'c,
and David Brin are also acknowledged.

\section*{References}
\refe Adams, F. C. and Laughlin, G. 1997, {\it Rev. Mod. Phys.}
{\bf 69}, 337.

\refe Annis, J. 1999, {\it J. Brit. Interplan. Soc.} {\bf 52}, 19
(preprint astro-ph/9901322).

\refe Balashov, Yu. V. 1994, {\it Studies in History and
Philosophy of Science\/} {\bf 25B}, 933.

\refe Ball, J. A. 1973, {\it Icarus\/} {\bf  19}, 347.

\refe Barrow, J. D. and Tipler, F. J. 1986, {\it The Anthropic
Cosmological Principle\/}  (Oxford University Press, New York).

\refe Bondi, H. and Gold, T. 1948, {\it Mon. Not. R. astr. Soc.}
{\bf 108}, 252.

\refe Bracewell, R. N. 1975 {\it The Galactic Club: Intelligent
Life in Outer Space\/} (W. H. Freeman, San Francisco).

\refe Carter, B. 1983, {\it Philos. Trans. R. Soc. London A} {\bf
310}, 347.

\refe \'Cirkovi\'c, M. M. 2003, {\it Am.\ J.\ Phys.}\ {\bf 71},
122.

\refe \'Cirkovi\'c, M. M. and Bostrom, N. 2000 {\it Astrophys.
Space Sci.} {\bf 274}, 675.

\refe Clarke, J. N. 1981, {\it Icarus\/} {\bf 46}, 94.

\refe Dyson, F. 1979, {\it Rev. Mod. Phys.} {\bf 51}, 447.

\refe Fogg, M. J. 1987, {\it Icarus\/} {\bf 69}, 370.

\refe Hart, M. H. 1975, {\it Q. Jl. R. astr. Soc.} {\bf 16}, 128.

\refe Hoyle, F. 1948, {\it Mon. Not. R. astr. Soc.} {\bf 108},
372.

\refe Kragh, H. 1996, {\it Cosmology and Controversy\/}
(Princeton University Press, Princeton).

\refe Krauss, L. M. and Chaboyer, B. 2003, {\it Science\/} {\bf
299}, 65.

\refe Lem, S. 1984, {\it His Master's Voice\/} (Harvest Books,
Fort Washington).

\refe Lem, S. 1987, {\it Fiasco\/} (Harcourt, New York).

\refe Lineweaver, C. H. 2001, {\it Icarus\/} {\bf 151}, 307.

\refe Norris, R. P. 2000, in {\it When SETI Succeeds: The impact
of high-information Contact}, ed. A. Tough (Foundation for the
Future, Washington DC).

\refe Olum, K. D. 2003, "Conflict between anthropic reasoning and
observation" (preprint gr-qc/0303070).

\refe Scalo, J. and Wheeler, J. C. 2002, {\it Astrophys. J.} {\bf
566}, 723.

\refe Shklovskii, I. S. and Sagan, C. 1966, {\it Intelligent Life
in the Universe\/} (Holden-Day, San Francisco).

\refe Tipler, F. J. 1980, {\it Q. Jl. R. astr. Soc.} {\bf 21},
267.

\refe Walker, M. A. and \'Cirkovi\'c, M. M. 2003, "The Fermi
Equation," preprint.

\refe Walters, C., Hoover, R. A., and Kotra, R. K. 1980, {\it
Icarus\/} {\bf 41}, 193-197.

\refe Ward, P. D. and Brownlee, D. 2000, {\it Rare Earth: Why
Complex Life Is Uncommon in the  Universe\/} (Springer, New York).

\refe Wilson, T. L. 1984, {\it Q. Jl. R. astr. Soc.} {\bf 25},
435-448.

\end{document}